\documentclass[preprint,onecolumn,nofootinbib]{revtex4}
\pdfoutput=1
\usepackage[colorlinks=true,linkcolor=blue,urlcolor=blue,filecolor=black,citecolor=red,pdfstartview=FitV,pdftitle={},pdfsubject={},pdfkeywords={},pdfpagemode=None,bookmarksopen=true]{hyperref}
\usepackage{graphicx}%Include figure files
\usepackage{mathrsfs}%包含命令\mathscr{}（花体字母）
\usepackage{amsmath}
\usepackage{amsfonts}
\usepackage{amssymb,ulem}
\usepackage{color,xcolor}%
\usepackage{CJK}
\usepackage{subfigure}

\usepackage{dcolumn}% Align table columns on decimal pointl

\newcommand{\f}{\begin{equation}}
\newcommand{\ff}{\end{equation}}
\newcommand{\fa}{\begin{eqnarray}}
\newcommand{\ffa}{\end{eqnarray}}
\begin{document}

\title{Gravitational waves with generalized holonomy corrections}

\author{Shulan Li$^{1}$}
\thanks{shulanli.yzu@gmail.com}
\author{Jian-Pin Wu$^{1}$}
\thanks{jianpinwu@yzu.edu.cn, corresponding author} 
\affiliation{
  $^1$ Center for Gravitation and Cosmology, College of Physical Science and Technology, Yangzhou University, Yangzhou 225009, China}
%  \\ $^2$ Institute of Theoretical Physics, School of Physics, Dalian University of Technology, Dalian 116024, China.}

%----------------------------------------------------
\begin{abstract}
The cosmological tensor perturbation equation with generalized holonomy corrections is derived in the framework of effective loop quantum gravity. This results in a generalized dispersion relation for gravitational waves, encompassing holonomy corrections. Furthermore, we conduct an examination of the constraint algebra concerning vector modes with generalized holonomy corrections. The requirement of anomaly cancellation for vector modes imposes constraints on the possible functional forms of the generalized holonomy corrections. What's more, we estimate the theoretical value of the effective graviton mass and discuss the potential detectability of this effective mass in future observations.
\end{abstract}
%----------------------------------------------------

\maketitle

\tableofcontents

%----------------------------------------------------
%----------------------------------------------------
\section{Introduction}
\label{introduction}
%----------------------------------------------------
Cosmological perturbation theory serves an important role in modern cosmology as the nexus between theory and observation, providing a fundamental framework and tool for investigating the large-scale structures of the universe, along with the origin and evolution of the anisotropy of the cosmic microwave background (CMB) radiation. Within the framework of cosmological perturbation theory, the universe's inhomogeneity and anisotropy are treated as fluctuations or perturbations of the spatially homogeneous and isotropic cosmic background. These equations of motion can be systematically solved order by order. Observations of the CMB radiation demonstrate that the universe only slightly deviates from this background on large scales with a magnitude of approximately $10^{-5}$. When studying such small fluctuations, the perturbation equations up to the linear order can already offer a sufficiently precise depiction of the early universe. The decomposition theorem \cite{Kodama:1984ziu, Mukhanov:1990me, kurki2005cosmological} states that the linear gravity perturbations can be decomposed into distinct modes, namely scalar, vector and tensor modes. This decomposition is based on the different transformation properties of these modes under spatial rotations. These modes evolve autonomously and remain uncoupled, allowing for independent analysis of distinct perturbation modes. In this paper, we will study the tensor and vector modes within the framework of effective loop quantum gravity (LQG) with generalized holonomy corrections.

In cosmology, delving into the study of gravitational waves (GWs) that emerged from the intricate physical processes of the early universe offers an unparalleled opportunity to peer into the universe's primordial stages.  These GWs, often referred to as tensor mode perturbations, carry with them an imprint of the conditions and events that prevailed during the universe's formative stages. Currently, considerable scientific endeavors are dedicated to detecting the subtle signatures left by these tensor mode perturbations on the structure of spacetime. This is achieved by measuring the polarization in the CMB. 

Given the potential significance of quantum gravity effects in the universe's early stage, there exists a compelling motivation to delve into the exploration of quantum gravity's possible impact on the propagation of GWs throughout these epochs. The study of such quantum gravity effects on the propagation of GWs holds the promise of shedding light on the fundamental nature of spacetime itself during these early stages.

Among various prospective quantum gravity theories, LQG, a nonperturbative and background-independent quantum gravity theory \cite{Rovelli:2004tv, Thiemann:2007pyv, Ashtekar:2004eh, Han:2005km}, has recently emerged as a prominent contender and attracted significant attention. The technique of LQG has been effectively employed to achieve quantization within the framework of symmetry-reduced cosmological spacetime, which is commonly known as loop quantum cosmology (LQC) \cite{Bojowald:2001xe, Ashtekar:2006rx, Ashtekar:2006uz, Ashtekar:2006wn, Bojowald:2005epg, Ashtekar:2003hd, Ashtekar:2011ni}. The effective theory of LQC can be formulated by incorporating two types of primary quantum gravitational effects: the inverse volume corrections and the holonomy corrections. These effects can be introduced through both the canonical approach \cite{Taveras:2008ke, Ding:2008tq, Yang:2009fp, Bojowald:2009jj, Bojowald:2009jk, Bojowald:2010qm} and the path integral perspectives \cite{Ashtekar:2009dn, Ashtekar:2010ve, Ashtekar:2010gz, Huang:2011es, Qin:2012gaa, Qin:2011hx, Qin:2012xh}.
Significant advancements have been achieved in the applications of LQC within the early universe in recent years. Notably, LQC has successfully replaced the classical big bang singularity from general relativity (GR) with a non-singular big bounce at the effective level, as demonstrated by key works such as Refs. \cite{Bojowald:2001xe, Ashtekar:2006rx, Ashtekar:2006uz}. It can also provide potential origins for the inflationary model \cite{Bojowald:2002nz} and set appropriate initial conditions \cite{Bojowald:2003mc}. Furthermore, quantum gravitational effects may also leave detectable imprints, which can be observed in the CMB \cite{Bojowald:2003mc,Tsujikawa:2003vr}. In summary, LQC is the symmetry-reduced model of LQG, thus inheriting its fundamental characteristics. Within the realm of LQG research, LQC plays a twofold role: firstly, it explains the physical properties of the universe, and secondly, it severs as a natural laboratory to test LQG.

In this paper, we delve into the investigation of LQG effects concerning tensor modes. Specifically, our focus lies in the exploration of the generalized holonomy corrections on the dynamics of GWs. An exemplary effective holonomy correction involves substituting the classical background connection variable, denoted as $\bar{k}$, with the holonomy function $\frac{\sin\left(\gamma\bar{\mu}\bar{k}\right)}{\gamma\bar{\mu}}$, where $\bar{\mu}$ signifies the polymerisation scale and $\gamma$ is the Barbero-Immirzi parameter \cite{BarberoG:1994eia, Immirzi:1996di}. In the classical limit where $\bar \mu  \to 0$, the effective LQG system recovers the classical counterpart. It's noteworthy that this substitution takes place due to our adoption of the minimal spin number $j=1/2$ in the SU(2) representation. Owing to the inherent quantization ambiguities, the explicit formulations of holonomy corrections are far from unique. For instance, diverse representations with different $j$ values have been investigated extensively in prior research works \cite{Gaul:2000ba,Vandersloot:2005kh,Chiou:2009hk,BenAchour:2016ajk}. Additionally, alternative formulations, incorporating quantization ambiguities or higher powers of $\bar{\mu}$, have been explored, as evidenced by references such as \cite{Hrycyna:2008yu,Yang:2009fp}.

Recently, the generalized holonomy corrections utilizing the substitution $\bar{k} \to g\left( {\bar k,\bar p} \right)$, where $\bar p$ is the conjugate momentum of $\bar k$ and $g\left( {\bar k,\bar p} \right)$ is a generalized undetermined function, have been put forth in the work of Ref. \cite{Han:2017wmt}. Through the adoption of the anomaly-free constraint algebra approach or the deformed algebra approach proposed in Refs. \cite{Bojowald:2008gz, Bojowald:2008jv}, and by strategically incorporating specific counter terms, they successfully achieve an anomaly-free constraint algebra. Notably, they are able to determine and address all the necessary counter terms within the constraints, thereby elucidating the restriction on this undetermined function $g\left( {\bar k,\bar p} \right)$. And then they work out the gauge invariant perturbation equations together with the corresponding inflationary power spectra. In a more recent study \cite{Renevey:2021tmh}, the authors employ a comprehensive investigation into the background dynamics of the inflationary phase, considering generalized holonomy corrections. Their research also delves into the impact of these corrections on the primordial scalar power spectrum. It is noteworthy, however, that they employ the classical scalar perturbation equation at the perturbative level, which may lead to a loss of information regarding the generalized holonomy corrections at the level of perturbations. Notably, in a subsequent work \cite{DeSousa:2022rep}, the authors explore the primordial power spectra incorporating generalized holonomy corrections based on perturbation equations derived within the framework of deformed algebra presented in Ref. \cite{Han:2017wmt}. This approach retains information about the generalized holonomy corrections at the perturbative level.

However, the inclusion of the counter terms within the constraints has the potential to introduce quantum gravity corrections that extend beyond the scope of holonomy corrections. It is undoubtedly important to focus solely the effects of holonomy corrections on GW dynamics. Therefore, we will exclude counter terms and consider only holonomy corrections, which may otherwise bring unsuspected corrections. Since we considered a generalized holonomy corrected function, which is undertermined, it is possible to achieve anomaly freedom in perturbative LQG, even though excluding the incorporation of counter terms. As a first step, we also study the anomaly-free constraint algebra of the vector modes in this work.

The organization of the paper is as follows. In Sec. \ref{1}, we recall the canonical formulation of GR and deduce its classical dynamics. In Sec. \ref{2}, the quantum-corrected tensor mode equation with generalized holonomy corrections along with the corresponding dispersion relation of GW propagation is derived. In Sec. \ref{3}, we consider anomaly cancellation in the constraint algebra for vector modes and obtain the restriction on the function $g\left( {\bar k,\bar p} \right)$. Besides, a specific example of the correction function $g\left( {\bar k,\bar p} \right)$ is introduced to provide a more intuitive and clear interpretation of the quantum dynamics and anomaly cancellation condition obtained earlier. In Sec. \ref{4}, we estimate the theoretical value of the effective graviton mass and  explore the potential detectability of this mass in future observations. In Sec. \ref{conclusion}, we summarize our results and conclusions, and then present our outlook. Throughout this paper, we adopt the Planck unit system, where $ c = G = \hbar = 1 $, with $c$ denoting the speed of light, $G$ representing the gravitational constant, and $\hbar$ signifying the reduced Planck constant.

%----------------------------------------------------
%----------------------------------------------------
\section{Canonical formulation and classical dynamics}
\label{1} %"the 1st section of the main body"
%----------------------------------------------------
In this section, following the idea in Ref. \cite{Bojowald:2007cd}, we will review the canonical formulation of GR and deduce its classical dynamics, which will help us derive the quantum dynamics with generalized holonomy corrections in the section that follows.

%----------------------------------------------------
\subsection{Spacetime metric in canonical formulation}
\label{1.1} %"1.1 of the main body"
%----------------------------------------------------
A spatially flat Friedmann-Robertson-Walker (FRW) background spacetime can be described by the following metric:
\begin{align}
\label{FRWBM} %"FRW Background Metric"
{\rm{d}}{s_{\rm{b}}}^2 &=  - {\rm{d}}{t^2} + {a^2}\left( t \right){\delta _{ab}}{\rm{d}}{x^a}{\rm{d}}{x^b}\nonumber\\
 &= {a^2}\left( \eta  \right)\left( { - {\rm{d}}{\eta ^2} + {\delta _{ab}}{\rm{d}}{x^a}{\rm{d}}{x^b}} \right)\;,
\end{align}
where $t$ denotes cosmic time, i.e. the proper time of isotropic observers, $\eta$ represents conformal time, and $a\left( t \right)$ or $a\left( \eta \right)$ is the scale factor. The indices $a,b,\dots$ denote spatial coordinates. When only taking into account the perturbations of the tensor modes, the metric can be expressed as follows:
\begin{align} 
\label{TPM} %"Tensor Perturbation Metric"
{\rm{d}}{s_{\rm{t}}}^2 &=  - {\rm{d}}{t^2} + {a^2}\left( t \right)\left( {{\delta _{ab}} + {h_{ab}}} \right){\rm{d}}{x^a}{\rm{d}}{x^b}\nonumber\\
 &= {a^2}\left( \eta  \right)\left[ { - {\rm{d}}{\eta ^2} + \left( {{\delta _{ab}} + {h_{ab}}} \right){\rm{d}}{x^a}{\rm{d}}{x^b}} \right]\;.
\end{align}
In this equation, the symmetric metric perturbation field $h_{ab}$ adheres to the conditions of being transverse and traceless, i.e. ${\partial^a}{h_{ab}} = {\delta^{ab}}{h_{ab}} = 0$, which effectively removes any vectorial or scalar contributions. In a canonical formulation based on the $3+1$ decomposition of spacetime, the metric can be expressed in terms of the spatial metric $q_{ab}$, the lapse function $N$, and the shift vector $N^a$ as follows:
\begin{align}
\label{CF} %"Canonical Formulation"
{\rm{d}}{s^2} &=  - {N^2}{\left( {{\rm{d}}{x^0}} \right)^2} + {q_{ab}}\left( {{\rm{d}}{x^a} + {N^a}{\rm{d}}{x^0}} \right)\left( {{\rm{d}}{x^b} + {N^b}{\rm{d}}{x^0}} \right)\nonumber\\
&= \left( { - {N^2} + {q_{ab}}{N^a}{N^b}} \right){\left( {{\rm{d}}{x^0}} \right)^2} + 2{q_{ab}}{N^b}{\rm{d}}{x^0}{\rm{d}}{x^a} + {q_{ab}}{\rm{d}}{x^a}{\rm{d}}{x^b}\;.
\end{align}
This canonical formulation serves as a powerful framework for effectively depicting the dynamics of spacetime.

%----------------------------------------------------
\subsection{Canonical variables in Ashtekar’s formulation}
\label{1.2} %"1.2 of the main body"
%----------------------------------------------------
Within the framework of LQG, GR is reformulated using Ashtekar's formulation \cite{Ashtekar1987New, BarberoG:1994eia}. In this formulation, the spatial metric $q_{ab}$ is replaced by the densitized triad $E_i^a$, as defined by:
\begin{equation}
	\label{Eia} %"the definition of E_i^a"
	E_i^a: = \left| {\det \left( {e_b^j} \right)} \right|e_i^a\;.
\end{equation}
Here, $e_i^a$ and $e_a^i$ represent the triad and its dual, respectively, satisfying ${q_{ab}} = e_a^i e_b^i$. The indices $i,j,\dots$ correspond to internal indices. 

Another important basic variable is the Ashtekar-Barbero connection:
\begin{equation} 
\label{Aai} %"the definition of A_a^i"
A_a^i: = \Gamma _a^i + \gamma K_a^i \;.
\end{equation}
Here, $K_a^i$ represents the extrinsic curvature, and the Barbero-Immirzi parameter $\gamma$ is a positive value, i.e., $\gamma > 0$ \cite{BarberoG:1994eia, Immirzi:1996di}. The spin connection $\Gamma _a^i$, which corresponds to the covariant derivative $D_a$ satisfying ${D_a}e_i^a = 0$ by definition, is determined as follows:
\begin{equation}
\label{Gamma ai} %"the expression of \Gamma_a^i"
\Gamma _a^i =  - {\epsilon ^{ijk}}e_j^b\left( {{\partial _{[a}}e_{b]}^k + \frac{1}{2}e_k^ce_a^l{\partial _{[c}}e_{b]}^l} \right)\;.
\end{equation}
The Ashtekar-Barbero connection $A_a^i$ is the canonically conjugate variable of the densitized triad $E_i^a$ and characterizes spatial curvature.

 After perturbing the spacetime, for any physical variable $V$, the background part and the perturbation part are respectively represented by $\bar V$ and $\delta V$, that is, $V = \bar V + \delta V$. By matching Eqs. (\ref{FRWBM}) and (\ref{CF}) and utilizing relevant definitions, we can derive the background variables as follows:
\begin{gather}
{{\bar q}_{ab}} = \bar p{\delta _{ab}}\;,\quad {{\bar N}^a} = 0\;,\quad \bar N = \left\{ {\begin{array}{*{20}{l}}
{1,}&{{x^0} = t}\\
{\sqrt {\bar p} ,}&{{x^0} = \eta }
\end{array}} \right.\;,\nonumber\\
\bar E_i^a = \bar p\delta _i^a\;,\quad \bar K_a^i = \bar k\delta _a^i\;,\quad \bar \Gamma _a^i = 0\;,
\label{BVs} %"Background Variables"
\end{gather}
where $\bar p = a^2$. 

By comparing Eqs. (\ref{TPM}) and (\ref{CF}), and leveraging the relevant definitions and background variables in Eq. (\ref{BVs}), the perturbative variables are computed at the linear level to yield:
\begin{gather}
\delta N = 0\;,\quad \delta {N^a} = 0\;,\nonumber\\
\delta E_i^a =  - \frac{1}{2}\bar ph_i^a\;,\quad \delta \Gamma _a^i = \frac{1}{{\bar p}}{\epsilon ^{ije}}{\delta _{ac}}{\partial _e}\delta E_j^c\;.
\label{PVs} %"Perturbation Variables"
\end{gather}
Regarding the extrinsic curvature $K_a^i$, determining the explicit expressions for both the background component $\bar k$ and the perturbation component $\delta K_a^i$ necessitates solving the Hamiltonian canonical equations, a process that will be elaborated on later.

Within the context of linear perturbed theory, the independent phase space variables are $\left( {\bar k,\bar p} \right)$ and $\left( {\delta K_a^i,\delta E_i^a} \right)$, whereas the symplectic structure is given by the following non-trivial Poisson brackets \cite{Bojowald:2006tm, Bojowald:2007hv, Bojowald:2007cd, Bojowald:2008gz, Bojowald:2008jv,Wu:2012mh,Wu:2018mhg}:
\begin{equation}
\label{PBs} %"Poisson Brackets"
\left\{ {\bar k,\bar p} \right\} = \frac{{8{\rm{\pi }}}}{{3{V_0}}}\;,\quad \left\{ {\delta K_a^i\left( x \right),\delta E_j^b\left( y \right)} \right\} = 8{\rm{\pi }}\delta _j^i\delta _a^b{\delta ^3}\left( {x - y} \right)\;.
\end{equation}
Here ${V_0}: = \int_\Sigma  {{{\rm{d}}^3}x}$ is a fiducial volume introduced to ensure a finite symplectic structure for the background variables. This is accomplished by confining the integration of the action to a finite cell, rather than extending it across the entirety of space $\mathbb{R}^3$. The restriction to a cell doesn't result in information loss due to the background's homogeneity. It's important to note that this fiducial quantity only appears in the basic variables and their symplectic structure, without influencing the ultimate physical outcomes. Consequently, there are separate canonical structures for the background and the perturbations, yet these variables will be dynamically coupled. The dynamical homogeneous background, in particular, would receive back-reaction effects beyond the linear order.

%----------------------------------------------------
\subsection{Classical dynamics}
\label{1.3} %"1.3 of the main body"
%----------------------------------------------------
Since GR is a fully constrained theory, its Hamiltonian is formulated as a summation of smeared constraints. In the context of a canonical triad formulation, these constraints encompass three categories: the Hamiltonian constraint, the diffeomorphism constraint, and the Gauss constraint. However, when one focuses on linear perturbations involving solely the tensor mode, only the Hamiltonian constraint remains \cite{Bojowald:2007cd}. 

In the framework of the connection dynamical formalism, the Hamiltonian constraint is expressed as:
\begin{align}
	\label{HCv0} %"Hamiltonian Constraint"
	{H_{\rm{G}}}\left[ N \right] = \frac{1}{{16{\rm{\pi }}}}\int_\Sigma  {{{\rm{d}}^3}xN\frac{{E_j^cE_k^d}}{{\sqrt {\left| {\det E} \right|} }}\left[ {{\epsilon _i}^{jk}F_{cd}^i - 2\left( {1 + {\gamma ^2}} \right)K_{[c}^jK_{d]}^k} \right]}\;.
\end{align}
By utilizing the expressions of the perturbed basic variables and the curvature $F_{ab}^i = {\partial _a}A_b^i - {\partial_b}A_a^i + {\epsilon^i}_{jk}A_a^jA_b^k$, one can perturb the Hamiltonian constraint up to the quadratic terms in perturbations, yielding:
\begin{align}
\label{HC} %"Hamiltonian Constraint"
{H_{\rm{G}}}\left[ N \right]
 &= \frac{1}{{16{\rm{\pi }}}}\int_\Sigma  {{{\rm{d}}^3}x\bar N\left[ { - 6{{\bar k}^2}\sqrt {\bar p}  - \frac{{{{\bar k}^2}}}{{2{{\bar p}^{3/2}}}}\left( {\delta E_j^c\delta E_k^d\delta _c^k\delta _d^j} \right) + \sqrt {\bar p} \left( {\delta K_c^j\delta K_d^k\delta _k^c\delta _j^d} \right)} \right.} \nonumber\\
&\quad \quad \quad \quad \quad \quad \quad \quad \ \left. { - \frac{{2\bar k}}{{\sqrt {\bar p} }}\left( {\delta E_j^c\delta K_c^j} \right) + \frac{1}{{{{\bar p}^{3/2}}}}\left( {{\delta _{cd}}{\delta ^{jk}}{\delta ^{ef}}{\partial _e}E_j^c{\partial _f}E_k^d} \right)} \right] \;.
\end{align} 
It should be mentioned here that the terms dependent on $\gamma$ are eliminated from the Hamiltonian constraint when employing the spin connection and taking into account the symmetry of the densitized triad and extrinsic curvature for the tensor mode. 

In the standard covariant formulation, the equations of motion can be derived by taking the variation of the action with respect to the metric. In a canonical formulation, however, the equations of motion governing any phase space function $f$ can be elegantly expressed as $\dot f = \left\{ {f,H\left[ N \right]} \right\}$, wherein $H\left[ N \right]$ is the total Hamiltonian constraint, encompassing both the gravitational component $H_{\rm{G}}\left[ N \right]$ and the matter component $H_{\rm{M}}\left[ N \right]$. When $f$ represents the fundamental variables, its equations of motion are just the Hamiltonian canonical equations. In this context, the dot signifies the derivative with respect to the coordinate time $x^0$, which depends on the chosen background lapse function $\bar N$. 

Subsequently, we will compute the Hamiltonian canonical equations for both the background variables and the perturbed variables, respectively. Using the expression of the Hamiltonian constraint \eqref{HC}, we yield the following equations of motion for the background variables:
\begin{align}
\label{kdot} %"the expression of kbar dot"
\dot {\bar k} &= \left\{ {\bar k,H\left[ N \right]} \right\} =  - \frac{{\bar N{{\bar k}^2}}}{{2\sqrt {\bar p} }} + \frac{{8{\rm{\pi }}}}{{3{V_0}}}\frac{{\partial {H_{\rm{M}}}\left[ N \right]}}{{\partial \bar p}}\;,\\
\label{pdot} %"the expression of pbar dot"
\dot {\bar p} &= \left\{ {\bar p,{H_{\rm{G}}}\left[ N \right]} \right\} = 2\bar k\bar N\sqrt {\bar p} \;.
\end{align}
Since the matter Hamiltonian constraint ${H_{\rm{M}}}\left[ N \right]$ is independent of the extrinsic curvature, Eq. \eqref{pdot} remains unaffected by the matter Hamiltonian constraint, in contrast to Eq. \eqref{kdot}. When choosing $\bar N = \sqrt {\bar p}$, where the coordinate time $x^0$ corresponds to the conformal time $\eta$, according to Eq. (\ref{pdot}) and the relation $\bar p = a^2$, one can obtain the background extrinsic curvature ${\bar k = \frac{{\dot {\bar p}}}{{2\bar p}} = \frac{{\dot a}}{a} = :\mathscr{H}}$, which is nothing more than the conformal Hubble parameter.

We proceed to calculate the Hamiltonian canonical equation for the perturbed densitized triad $\delta E_i^a$, which is derived as: 
\begin{equation}
\label{dEdot} %"the expression of deltaE dot"
\delta \dot E_i^a = \left\{ {\delta E_i^a,{H_{\rm{G}}}\left[ N \right]} \right\} =  - \frac{1}{2}\bar N\sqrt {\bar p} \bar kh_i^a - \bar N\sqrt {\bar p} \delta _k^a\delta _i^c\delta K_c^k\;.
\end{equation}
While deriving the aforementioned equation, we have employed the expression of $\delta E_i^a$ as provided in Eq. (\ref{PVs}).

By combining Eq. \eqref{dEdot} with the time derivative of the expression of $\delta E_i^a$ in Eq. \eqref{PVs} and using Eq. \eqref{pdot}, the expression for the perturbed extrinsic curvature can be formulated as follows:
\begin{equation}
\label{dK} %"the expression of deltaK"
\delta K_a^i = \frac{1}{2}\frac{{\sqrt {\bar p} }}{{\bar N}}\dot h_a^i + \frac{1}{2}\bar kh_a^i\;.
\end{equation}
Then, using the $\delta E_i^a$ expression from Eq. \eqref{PVs} along with Eq. (\ref{dK}), we derive the Hamiltonian canonical equation for the perturbed extrinsic curvature $\delta K_a^i$ as follows:
\begin{equation}
\label{dKdot} %"the expression of deltaK dot"
\delta \dot K_a^i = \left\{ {\delta K_a^i,H\left[ N \right]} \right\} = 8{\rm{\pi }}\frac{{\delta {H_{\rm{M}}}\left[ N \right]}}{{\delta \left( {\delta E_i^a} \right)}} + \frac{1}{2}\frac{{\bar N}}{{\sqrt {\bar p} }}{\nabla ^2}h_a^i - \frac{1}{2}\bar k\dot h_a^i - \frac{1}{4}\frac{{\bar N}}{{\sqrt {\bar p} }}{\bar k^2}h_a^i\;.
\end{equation}
Moreover, by amalgamating Eq. (\ref{dKdot}) with the time derivative of Eq. (\ref{dK}) and incorporating Eqs. (\ref{kdot}) and (\ref{pdot}), one can deduce the equation governing the motion of tensor mode perturbations:
\begin{eqnarray}
\label{EoM} %"Equation of Motion"
\frac{1}{2}\left[ {\frac{{\sqrt {\bar p} }}{{\bar N}}\ddot h_a^i + \left( { - \frac{{\dot {\bar N}\sqrt {\bar p} }}{{{{\bar N}^2}}} + 3\bar k} \right)\dot h_a^i - \frac{{\bar N}}{{\sqrt {\bar p} }}{\nabla ^2}h_a^i} \right] = 8{\rm{\pi }}\Pi _a^i\;,
\end{eqnarray}
where
\begin{equation}
\label{Pi ai} %"the expression of \Pi_a^i"
\Pi _a^i = \frac{1}{{3{V_0}}}\frac{{\partial {H_{\rm{M}}}\left[ N \right]}}{{\partial \bar p}}\left( {\frac{{\delta E_j^c\delta _a^j\delta _c^i}}{{\bar p}}} \right) + \frac{{\delta {H_{\rm{M}}}\left[ N \right]}}{{\delta \left( {\delta E_i^a} \right)}}\;.
\end{equation}
The quantity $\Pi _a^i$ characterizes linear source terms that are both transverse and traceless, and these terms can be associated with the transverse and traceless component of the perturbed stress-energy tensor through the relation $\Pi _a^i = \bar p\delta {T^{\left( {{\rm{TT}}} \right)}}_a^i$. For an in-depth discussion, please refer to Ref. \cite{Bojowald:2007cd}.

Furthermore, the choice of $\bar N = \sqrt {\bar p}$ results in Eq. (\ref{EoM}) transforming into:
\begin{equation}
\label{EoM1} %"Equation of Motion of form 1"
\frac{1}{2}\left( {\ddot h_a^i + 2\bar k\dot h_a^i - {\nabla ^2}h_a^i} \right) = 8{\rm{\pi }}\Pi _a^i\;.
\end{equation}
Alternatively, setting $\bar N = 1$ to correspond with cosmic time $t$ yields Eq. (\ref{EoM}) taking the form of:
\begin{equation}
\label{EoM2} %"Equation of Motion of form 2"
\frac{1}{2}\left( {\sqrt {\bar p} \ddot h_a^i + 3\bar k\dot h_a^i - \frac{1}{{\sqrt {\bar p} }}{\nabla ^2}h_a^i} \right) = 8{\rm{\pi }}\Pi _a^i\;.
\end{equation}

When the source terms $\Pi _a^i$ are absent, the aforementioned tensor mode equation demonstrates propagating wave solutions recognized as GWs in the given cosmological background. For the investigation of wave propagation, it is often advantageous to deduce the associated dispersion relation, illustrating the interrelation between the frequency $\omega$ and the proper wave vector $\vec{k}/a$. Notice that here the frequency $\omega$ corresponds to cosmic time $t$, and accordingly, the tensor perturbation equation of the version in Eq. (\ref{EoM2}) is to be applied. As our main interest is on local propagation taking no account of cosmic scales, we ignore the friction term, which is the first-order time derivative term originating from cosmological expansion and proportional to $\bar k$ and, consequently, to the Hubble parameter.

By employing a plane wave ansatz $h_a^i \propto {\rm{exp}}\left( {{\rm{i}}\omega t - {\rm{i}}\vec{k} \cdot \vec{x}} \right)$, one arrives at:
\begin{equation}
	\label{DR} %"Dispersion Relation"
	{\omega ^2} = {\left( {\frac{k}{a}} \right)^2}\;,
\end{equation}
which is the well-known classical dispersion relation. This relation gives rise to the group velocity of GWs as follows:
\begin{equation}
	\label{GVoGWs} %"Group Velocity of GWs"
	{v_{{\rm{GW}}}}: = \frac{{{\rm{d}}\omega }}{{{\rm{d}}\left( {{k/a}} \right)}} = 1\;.
\end{equation}
As expected, the group velocity of classical GWs is exactly equal to the speed of light.

%----------------------------------------------------
%----------------------------------------------------
\section{Quantum dynamics with generalized holonomy corrections}
\label{2} %"the 2nd section of the main body"
%----------------------------------------------------
\subsection{Corrected tensor mode equation}
\label{2.1} %"2.1 of the main body"
%----------------------------------------------------
So far, the explicit form of quantum corrections in LQC lacks uniqueness and subjects to quantization ambiguities. Building upon the idea described in Ref. \cite{Han:2017wmt}, in order to introduce holonomy corrections, we introduce a generalized undefined function $g\left( {\bar k,\bar p} \right)$. In the classical limit, $g\left( {\bar k,\bar p} \right)$ is required to satisfy $g\left( {\bar k,\bar p} \right) \to \bar k$, thereby ensuring that the standard GR is recovered.
After replacing $\bar k \to g\left( {\bar k,\bar p} \right)$, the classical Hamiltonian constraint \eqref{HC} becomes:
\begin{align}
\label{QHC} %"Quantum corrected Hamiltonian Constraint"
H_{\rm{G}}^{\rm{Q}}\left[ N \right] =& \frac{1}{{16{\rm{\pi }}}}\int_\Sigma  {{{\rm{d}}^3}x\bar N\left[ { - 6\sqrt {\bar p} {g^2}\left( {\bar k,\bar p} \right) - \frac{1}{{2{{\bar p}^{3/2}}}}{g^2}\left( {\bar k,\bar p} \right)\left( {\delta E_j^c\delta E_k^d\delta _c^k\delta _d^j} \right)} \right.} \nonumber\\
&\left. { + \sqrt {\bar p} \left( {\delta K_c^j\delta K_d^k\delta _k^c\delta _j^d} \right) - \frac{2}{{\sqrt {\bar p} }}g\left( {\bar k,\bar p} \right)\left( {\delta E_j^c\delta K_c^j} \right) + \frac{1}{{{{\bar p}^{3/2}}}}\left( {{\delta _{cd}}{\delta ^{jk}}{\delta ^{ef}}{\partial _e}E_j^c{\partial _f}E_k^d} \right)} \right] \;.
\end{align}

We would like to emphasize that, in our approach, the background component $\bar{k}$ is substituted by the same quantum correction function $g\left( {\bar k,\bar p} \right)$, regardless of whether it pertains to background or perturbation terms. This is the most reasonable consideration. However, when we consider the conventional holonomy corrections as $\bar{k}\rightarrow \frac{{\sin \left(\gamma \bar \mu \bar k\right)}}{{\gamma \bar \mu }}$, we encounter issues with achieving an anomaly-free constraint algebra, even when applying it exclusively to vector modes, if the parameter $m$ assumes identical values in both background and perturbation terms \cite{Bojowald:2007cd, Bojowald:2007hv}. In the subsequent section, we will demonstrate that by employing the aforementioned generalized holonomy corrections, we can achieve an anomaly-free constraint algebra in the context of vector modes.

Following the same procedure as in the classical case discussed in subsection \ref{1.3}, we will derive the tensor mode equation with generalized holonomy corrections based on the quantum-corrected Hamiltonian constraint (\ref{QHC}).
Firstly, we calculate the Hamiltonian canonical equations of $\bar k$, $\bar p$ and $\delta E_i^a$, which are given as follows:
\begin{align}
\label{Qkdot} %"Quantum corrected expression of kbar dot"
\dot {\bar k} &= \left\{ {\bar k,H_{\rm{G}}^{\rm{Q}}\left[ N \right] + {H_{\rm{M}}}\left[ N \right]} \right\} = \frac{{8{\rm{\pi }}}}{{3{V_0}}}\frac{{\partial {H_{\rm{M}}}\left[ N \right]}}{{\partial \bar p}} - \frac{{\bar N}}{{2\sqrt {\bar p} }}{g^2}\left( {\bar k,\bar p} \right) - \bar N\sqrt {\bar p} \frac{\partial }{{\partial \bar p}}{g^2}\left( {\bar k,\bar p} \right)\;,\\
\label{Qpdot} %"Quantum corrected expression of pbar dot"
\dot {\bar p} &= \left\{ {\bar p,H_{\rm{G}}^{\rm{Q}}\left[ N \right]} \right\} = \bar N\sqrt {\bar p} \frac{\partial }{{\partial \bar k}}{g^2}\left( {\bar k,\bar p} \right) \;,\\
\label{QdEdot} %"Quantum corrected expression of deltaE dot"
\delta \dot E_i^a &= \left\{ {\delta E_i^a,H_{\rm{G}}^{\rm{Q}}\left[ N \right]} \right\} =  - \frac{1}{2}\bar N\sqrt {\bar p} h_i^ag\left( {\bar k,\bar p} \right) - \bar N\sqrt {\bar p} \delta _k^a\delta _i^c\delta K_c^k \;.
\end{align}

It is necessary to mention that the expression of $\delta E_i^a$ in Eq. (\ref{PVs}) still remains applicable in the present scenario with quantum corrections. By combining the results from the aforementioned Hamiltonian canonical equations and the time derivative of the expression of $\delta E_i^a$ in Eq. (\ref{PVs}), we arrive at the modified formulation for perturbed extrinsic curvature:
\begin{equation}
\label{QdK} %"Quantum corrected expression of deltaK"
\delta K_a^i = \frac{1}{2}\frac{{\sqrt {\bar p} }}{{\bar N}}\dot h_a^i + \frac{1}{2}h_a^i\frac{\partial }{{\partial \bar k}}{g^2}\left( {\bar k,\bar p} \right) - \frac{1}{2}h_a^ig\left( {\bar k,\bar p} \right)\;.
\end{equation}
Then the Hamiltonian canonical equation of $\delta K_a^i$ can be computed as:
\begin{align}
\label{QdKdot} %"Quantum corrected expression of deltaK dot"
&\delta \dot K_a^i = \left\{ {\delta K_a^i,H_{\rm{G}}^{\rm{Q}}\left[ N \right] + {H_{\rm{M}}}\left[ N \right]} \right\} \nonumber\\
&= 8{\rm{\pi }}\frac{{\delta {H_{\rm{M}}}\left[ N \right]}}{{\delta \left( {\delta E_i^a} \right)}} + \frac{1}{2}\frac{{\bar N}}{{\sqrt {\bar p} }}{\nabla ^2}h_a^i - \frac{1}{2}\dot h_a^ig\left( {\bar k,\bar p} \right) + \frac{{\bar N}}{{\sqrt {\bar p} }}h_a^i\left[ {\frac{3}{4}{g^2}\left( {\bar k,\bar p} \right) - \frac{1}{2}g\left( {\bar k,\bar p} \right)\frac{\partial }{{\partial \bar k}}{g^2}\left( {\bar k,\bar p} \right)} \right] \;.
\end{align}

By performing a time derivative on Eq. \eqref{QdK} and then comparing it with Eq. \eqref{QdKdot}, we can derive the tensor mode equation incorporating generalized holonomy corrections:
\begin{equation}
\label{QEoM} %"Quantum corrected Equation of Motion"
\frac{1}{2}\left\{ {\frac{{\sqrt {\bar p} }}{{\bar N}}\ddot h_a^i + \left[ { - \frac{{\dot {\bar N}\sqrt {\bar p} }}{{{{\bar N}^2}}} + \frac{3}{2}\frac{\partial }{{\partial \bar k}}{g^2}\left( {\bar k,\bar p} \right)} \right]\dot h_a^i - \frac{{\bar N}}{{\sqrt {\bar p} }}{\nabla ^2}h_a^i + {\frac{{\bar N}}{{\sqrt {\bar p} }}}{T_{\rm{Q}}}h_a^i} \right\} = 8{\rm{\pi }}{\Pi _{\rm{Q}}}_a^i \;,
\end{equation}
where
\begin{align}
\label{TQ} %"the expression of T_Q"
{T_{\rm{Q}}} =&  - \frac{3}{2}{g^2}\left( {\bar k,\bar p} \right) + g\left( {\bar k,\bar p} \right)\frac{\partial }{{\partial \bar k}}{g^2}\left( {\bar k,\bar p} \right)\nonumber\\
&  - \bar p\frac{\partial }{{\partial \bar k}}{g^2}\left( {\bar k,\bar p} \right)\frac{\partial }{{\partial \bar p}}g\left( {\bar k,\bar p} \right) + \frac{1}{2}{g^2}\left( {\bar k,\bar p} \right)\frac{\partial }{{\partial \bar k}}g\left( {\bar k,\bar p} \right)\nonumber\\
&  + \bar p\frac{\partial }{{\partial \bar k}}g\left( {\bar k,\bar p} \right)\frac{\partial }{{\partial \bar p}}{g^2}\left( {\bar k,\bar p} \right) + \bar p\frac{\partial }{{\partial \bar k}}{g^2}\left( {\bar k,\bar p} \right)\frac{\partial }{{\partial \bar k}}\frac{\partial }{{\partial \bar p}}{g^2}\left( {\bar k,\bar p} \right)\nonumber\\
&  - \frac{1}{2}{g^2}\left( {\bar k,\bar p} \right)\frac{{{\partial ^2}}}{{\partial {{\bar k}^2}}}{g^2}\left( {\bar k,\bar p} \right) - \bar p\frac{{{\partial ^2}}}{{\partial {{\bar k}^2}}}{g^2}\left( {\bar k,\bar p} \right)\frac{\partial }{{\partial \bar p}}{g^2}\left( {\bar k,\bar p} \right)\;,
\end{align}
and
\begin{equation}
\label{PiQ ai} %"the expression of \Pi_Q_a^i"
{\Pi _{\rm{Q}}}_a^i = \frac{1}{{3{V_0}}}\frac{{\partial {H_{\rm{M}}}\left[ N \right]}}{{\partial \bar p}}\left( {\frac{{\delta E_j^c\delta _a^j\delta _c^i}}{{\bar p}}} \right)\left[ { - \frac{\partial }{{\partial \bar k}}g\left( {\bar k,\bar p} \right) + \frac{{{\partial ^2}}}{{\partial {{\bar k}^2}}}{g^2}\left( {\bar k,\bar p} \right)} \right] + \frac{{\delta {H_{\rm{M}}}\left[ N \right]}}{{\delta \left( {\delta E_i^a} \right)}} \;.
\end{equation}
%It is worth mentioning that we completed the cumbersome calculations in the above derivations as well as that in subsequent contents with the aid of the Mathematica package xAct \cite{xAct}.

As expected, both the friction term and the source terms arising from the matter Hamiltonian receive the generalized holonomy corrections. Moreover, an additional term emerges, proportionally related to the field perturbation $h_a^i$. It's worth highlighting that the nature of these corrections is intricately linked to the specific functional form of the correction function $g\left( {\bar k,\bar p} \right)$.

In conclusion, a generalized correction function $g\left( {\bar k,\bar p} \right)$ has the potential to bridge the theoretical framework with empirical observations. Furthermore, the emergence of the extra term, linked to the field perturbation $h_a^i$, adds a layer of complexity to the corrected equations. This term's significance and implications warrant a comprehensive analysis. Its existence might entail novel physical interpretations or provide insights into previously uncharted aspects of the phenomena being studied.

%----------------------------------------------------
\subsection{Modified dispersion relation}
\label{2.2} %"2.2 of the main body"
%----------------------------------------------------
In this subsection, we proceed to study the modified dispersion relation of GWs resulting from the inclusion of generalized holonomy corrections. Only the additional term proportionally related to the field perturbation $h_a^i$ receives the holonomy corrections once the source terms are removed and the friction term is ignored. By performing the computations of the classical scenario but employing the holonomy corrected wave equation (Eq. \eqref{QEoM}), we arrive at the dispersion relation with generalized holonomy corrections:
\begin{equation}
\label{QDR} %"Quantum corrected Dispersion Relation"
{\omega ^2} = {\left( {\frac{k}{a}} \right)^2} + {m_{\rm{G}}}^2 \;,
\end{equation}
where
\begin{equation}
\label{mG} %"the definition of m_G ^2"
{m_{\rm{G}}}^2: = \frac{{{T_{\rm{Q}}}}}{{{a^2}}} \equiv \frac{{{T_{\rm{Q}}}}}{{\bar p}} \;.
\end{equation}

\begin{figure*}[ht]
	\centering
	\includegraphics[width=0.5\textwidth]{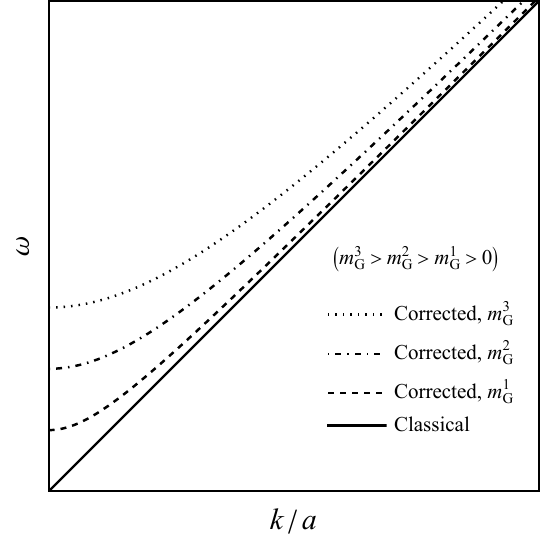}
	\caption{Dispersion relation for GWs with generalized holonomy corrections.}
	\label{fig1}
\end{figure*}

The modified dispersion relation (Eq. \eqref{QDR}) reveals that under quantum corrections, generalized holonomy corrections introduce a novel additive term ${m_{\rm{G}}}^2$ within the dispersion relation, as compared to the classical scenario described by equation (\ref{DR}). This signifies that GWs subject to these generalized holonomy corrections embody an effective mass ${m_{\rm{G}}}$. Remarkably, this modified dispersion relation (\ref{QDR}), regardless of the specific form of correction function $g\left( {\bar k,\bar p} \right)$, is formally consistent with that with the conventional holonomy corrections in Ref. \cite{Bojowald:2007cd}. However, the entire impact arising from distinct specific choices of the correction function $g\left( {\bar k,\bar p} \right)$ becomes encapsulated within the effective mass squared term ${m_{\rm{G}}}^2$. For visualization purposes, we present the modified dispersion relation with different ${m_{\rm{G}}}^2$ in Fig. \ref{fig1}.

The group velocity of GWs corresponding to the modified dispersion relation (\ref{QDR}) becomes:
\begin{equation}
\label{QGVoGWs} %"Quantum corrected Group Velocity of GWs"
v_{{\rm{GW}}}^{\rm{Q}}: = \frac{{{\rm{d}}\omega }}{{{\rm{d}}\left( {k/a} \right)}} = \frac{{\left( {k/a} \right)}}{{\sqrt {{{\left( {k/a} \right)}^2} + {m_{\rm{G}}}^2} }} \;.
\end{equation}
This indicates that different modes of GWs propagate with distinct group velocities that remain below unity. As a result, causality is maintained, provided that the condition of positivity holds for the effective mass squared term ${m_{\rm{G}}}^2$.

Obviously, the dependence of the group velocities $v_{{\rm{GW}}}^{\rm{Q}}$ on the effective mass ${m_{\rm{G}}}$ of GWs is evident, contingent upon the specific form of the generalized holonomy-correction function $g\left( {\bar k,\bar p} \right)$. When considering a fixed proper wave number $k/a$, the behavior of $v_{{\rm{GW}}}^{\rm{Q}}$ becomes apparent: it diminishes as ${m_{\rm{G}}}$ increases. This behavior aligns with expectations, as greater effective mass ${m_{\rm{G}}}$ corresponds to more pronounced quantum corrections received by the GWs.
Upon closer examination of GWs with identical effective mass ${m_{\rm{G}}}$, a notable observation arises: distinct $k/a$ modes of GWs exhibit varying group velocities $v_{{\rm{GW}}}^{\rm{Q}}$, consistently remaining below unity. Notably, the group velocities of lower $k/a$ modes deviate significantly from the classical case, while those with higher $k/a$ values gradually approach unity—the speed of light. This finding indicates that, for modes featuring large proper wave numbers $k/a$, the impact of holonomy corrections becomes negligible.

%----------------------------------------------------
%----------------------------------------------------
\section{Anomaly freedom of the vector modes}
\label{3} %"the 3rd section of the main body"
%----------------------------------------------------
We've successfully derived the GW equation with generalized holonomy corrections and discussed its associated dispersion relation. Moving forward, our focus will shift to addressing the constraint algebra for vector modes, along with investigating a particular instance of the generalized holonomy-correction function $g\left( {\bar k,\bar p} \right)$. Specifically, we will delve into the anomaly cancellation condition of the constraint algebra for vector modes and the parameter restrictions on this specific form of $g\left( {\bar k,\bar p} \right)$.

%----------------------------------------------------
\subsection{Anomaly cancellation in the constraint algebra for vector modes}
\label{3.1} %"3.1 of the main body"
%----------------------------------------------------
In the classical framework of GR, the Poisson brackets among the smeared constraints weakly vanish on the constraint surfaces, thereby forming a first-class constraint algebra. However, when quantum corrections are factored in, to ensure a consistent theory, it becomes imperative to maintain closedness within this algebraic structure. In the context of background dynamics alone, it is relatively straightforward to observe that the quantum-corrected constraints continue to yield a closed algebra.
Nevertheless, the scenario evolves as we include perturbations into the constraints. In this case, the resultant constraint algebra often deviates from closure, featuring extra terms referred to as anomaly terms. Achieving a well-defined, closed algebra necessitates the vanish of these anomaly terms.

Regarding the tensor mode with generalized holonomy corrections under consideration, as mentioned previously, only the Hamiltonian constraint remains. Its effective form $H_{\rm{G}}^{\rm{Q}}\left[ N \right]$ has been presented in Eq. (\ref{QHC}). It's evident that the Poisson bracket between two effective Hamiltonian constraints $\left\{ {H_{\rm{G}}^{\rm{Q}}\left[ {{N_1}} \right],H_{\rm{G}}^{\rm{Q}}\left[ {{N_2}} \right]} \right\}$ is identically zero due to the absence of lapse perturbations $\delta N$, aligning with classical results. Thus, the constraint algebra for the tensor mode are automatically anomaly-free.

Let us now shift our focus to examining the constraint algebra for vector modes. The gravitational contribution of the perturbed Hamiltonian constraint, considering only vector modes and accounting for generalized holonomy corrections, is expressed as follows:
\begin{align}
\label{vQHC} %"vector mode Quantum corrected Hamiltonian Constraint"
H_{\rm{vG}}^{\rm{Q}}\left[ N \right] =\frac{1}{{16{\rm{\pi }}}}&\int_\Sigma  {{{\rm{d}}^3}x\bar N \left[ { - 6\sqrt {\bar p} {g^2}\left( {\bar k,\bar p} \right) - \frac{1}{{2{{\bar p}^{3/2}}}}{g^2}\left( {\bar k,\bar p} \right)\left( {\delta E_j^c\delta E_k^d\delta _c^k\delta _d^j} \right)} \right.} \nonumber\\
&\quad\quad\quad\quad\ \left. { + \sqrt {\bar p} \left( {\delta K_c^j\delta K_d^k\delta _k^c\delta _j^d} \right) - \frac{2}{{\sqrt {\bar p} }}g\left( {\bar k,\bar p} \right)\left( {\delta E_j^c\delta K_c^j} \right)} \right] \;.
\end{align}
On the other hand, the perturbed gravitational diffeomorphism constraint, considered exempt from quantum corrections due to its direct quantization through phase space transformations \cite{Ashtekar:1995zh}, is formulated as follows:
\begin{equation}
\label{vDC} %"vector mode Diffeomorphism Constraint"
{D_{{\rm{vG}}}}\left[ {{N^a}} \right] = \frac{1}{{8{\rm{\pi }}}}\int_\Sigma  {{{\rm{d}}^3}x\;\delta {N^c}\left[ { - \bar p\left( {{\partial _k}\delta K_c^k} \right) - \bar k\delta _c^k\left( {{\partial _d}\delta E_k^d} \right)} \right]} \;.
\end{equation}

Once again, it's straightforward to observe that the Poisson bracket $\left\{ {H_{\rm{vG}}^{\rm{Q}}\left[ {{N_1}} \right],H_{\rm{vG}}^{\rm{Q}}\left[ {{N_2}} \right]} \right\}$ is trivial. However, the non-trivial Poisson bracket between $H_{\rm{vG}}^{\rm{Q}}\left[ N \right]$ and ${D_{{\rm{vG}}}}\left[ {{N^a}} \right]$ turns out to be:
\begin{align}
\label{PBvHD} %"Poisson Bracket between vector mode Hamiltonian constraint and Diffeomorphism constraint"
\left\{ {H_{{\rm{vG}}}^{\rm{Q}}\left[ N \right],{D_{{\rm{vG}}}}\left[ {{N^a}} \right]} \right\} = &\frac{{\bar N}}{{\sqrt {\bar p} }}\left[ {\bar k + g\left( {\bar k,\bar p} \right) - \frac{\partial }{{\partial \bar k}}{g^2}\left( {\bar k,\bar p} \right)} \right]{D_{\rm{vG}}}\left[ {{N^a}} \right] \nonumber\\
&+ \frac{1}{{8{\rm{\pi }}}}\int_\Sigma  {{{\rm{d}}^3}z\;} \bar p\left( {{\partial _c}\delta {N^j}} \right){\cal A}_j^c \;,
\end{align}
where
\begin{equation}
\label{Ajc} %"the expression of A_j^c"
{\cal A}_j^c = \frac{{\bar N}}{{\sqrt {\bar p} }}\left[ {\bar p\frac{\partial }{{\partial \bar p}}{g^2}\left( {\bar k,\bar p} \right) + {g^2}\left( {\bar k,\bar p} \right) - {{\bar k}^2} + \bar k\frac{\partial }{{\partial \bar k}}{g^2}\left( {\bar k,\bar p} \right) - 2\bar kg\left( {\bar k,\bar p} \right)} \right]\left( {\frac{{\delta E_j^c}}{{\bar p}}} \right) \;.
\end{equation}
The anomaly cancellation condition ${\cal A}_j^c = 0$ imposes constraints on the functional form of $g\left( {\bar k,\bar p} \right)$, leading to a first-order quadratic partial differential equation:
\begin{equation}
\label{PDE} %"Partial Differential Equation"
\bar p\frac{\partial }{{\partial \bar p}}{g^2}\left( {\bar k,\bar p} \right) + {g^2}\left( {\bar k,\bar p} \right) - {\bar k^2} + \bar k\frac{\partial }{{\partial \bar k}}{g^2}\left( {\bar k,\bar p} \right) - 2\bar kg\left( {\bar k,\bar p} \right) = 0 \;.
\end{equation}
Because of being a nonlinear partial differential equation, obtaining an explicit analytical general solution for Eq. (\ref{PDE}) remains challenging. Subsequently, our focus will shift towards examining a specific instance of the generalized holonomy-correction function $g\left( {\bar k,\bar p} \right)$.

%----------------------------------------------------
\subsection{Parameter restrictions for a specific correction function}
\label{3.2} %"3.2 of the main body"
%----------------------------------------------------
In Ref. \cite{Renevey:2021tmh}, the authors propose a particular expression for $g\left( {\bar k,\bar p} \right)$, given by:
\begin{equation}
\label{SFog} %"Specific Form of g()"
g\left( {\bar k,\bar p} \right) = \frac{{f\left( {\gamma \bar \mu \bar k} \right)}}{{\gamma \bar \mu }}\;,
\end{equation}
where $\bar \mu \left( {\bar p} \right) = \sqrt {\Delta/{\bar p}}$ following the standard $\bar{\mu}$-scheme. Here, $\Delta = 4\sqrt{3}\pi\gamma {l_{\text{P}}}^2$, where ${l_{\text{P}}}$ represents the Planck length—the fundamental length unit in the Planck unit system, in which it equals unity. This quantity serves as the LQG area gap, signifying the minimum non-zero eigenvalue of the quantum area operator. The function $f$ takes the form:
\begin{equation}
\label{fX-1} %"the 1st version of the expression of f(X)"
f\left( X \right) = \sqrt {\frac{{{{\left( {1 + CX} \right)}^{1 - \alpha }}}}{{1 + C}}\left( {1 - {{\cos }^{2\left( {C + 1} \right)}}\left( X \right)} \right)}\;.
\end{equation}
Here, both $\alpha$ and $C$ are undetermined parameters, with the specific note that $C$ is a non-negative integer. The independent variable of this function $f$ is only considered to range over $\left [ 0,\pi \right ]$ as indicated in Ref. \cite{Renevey:2021tmh}.

The provided function form $g\left( {\bar k,\bar p} \right)$ satisfies the prerequisite that $g\left( {\bar k,\bar p} \right) \to \bar k$ in the classical limit ($\bar \mu \to 0$). Furthermore, it's evident that by setting the parameter $C$ to zero, Eq. (\ref{SFog}) restores the standard $\bar{\mu}$-scheme of the holonomy corrections, i.e., $g\left( {\bar k,\bar p} \right) = \frac{{\sin \left(\gamma \bar \mu \bar k\right)}}{{\gamma \bar \mu }}$. Additional details and discussions regarding this specific form of $g\left( {\bar k,\bar p} \right)$ can be found in Ref. \cite{Renevey:2021tmh}.
This function form of $g\left( {\bar k,\bar p} \right)$ can serve as a simplified toy model to provide greater clarity and intuition for the quantum-corrected dynamics explored in Sec. \ref{2}, as well as for comprehending the constraints on the function form of $g\left( {\bar k,\bar p} \right)$ arising from the anomaly cancellation of vector modes, as discussed in the previous subsection. Moving forward, we will delve into the analysis of constraints on the free parameters $\alpha$ and $C$ in Eq. (\ref{fX-1}) based on the previously obtained results.

We would like to underscore that, despite adopting the specific form of the function $f$ as outlined in Eq. \eqref{fX-1}, obtaining an explicit analytical general solution for the anomaly cancellation condition \eqref{PDE} remains challenging. Fortunately, our main focus is on the propagation of GWs within the low-curvature phase, which is far away from the high-curvature stage of the very early universe. In this context, the background extrinsic curvature $\bar{k}$ (and therefore the Hubble parameter) is considered to be much lower than ${\cal O}\left( 1 \right)$. This means that the argument ${\gamma \bar \mu \bar k}$ of the function $f$ in Eq. \eqref{SFog} satisfies ${\gamma \bar \mu \bar k} \ll {\cal O}\left( 1 \right)$, taking account of $\gamma \sim {\cal O}\left( 1 \right)$ together with $\bar \mu \left( {\bar p} \right) = \sqrt {\Delta/{\bar p}} \equiv \sqrt{\Delta}/a \sim {\cal O}\left( 1 \right)$. The latter condition results from $\Delta \sim {\cal O}\left( 1 \right){l_{{\rm{P}}}}^2 \equiv {\cal O}\left( 1 \right)$ and $a\left( t \right) \sim {\cal O}\left( 1 \right)$, with the scale factor of the present universe set to $a\left( t_0 \right) = 1$. Therefore, in the subsequent analysis, we will examine the constraints on the parameters $\alpha$ and $C$ specifically in the low-curvature stage, where $\bar k \ll {\cal O}\left( 1 \right)$.

Substituting Eq. (\ref{SFog}) and Eq. (\ref{fX-1}) into Eq. (\ref{PDE}) resulting from the anomaly cancellation of vector modes, gives rise to the following equation:
\begin{equation}
\label{PR} %"Parameter Restriction"
\frac{5}{2}C\left( {1 - \alpha } \right)\gamma \bar \mu {\bar k^3} + {\cal O}\left( {{{\bar k}^4}} \right) = 0\;.
\end{equation}
This implies that the anomaly cancellation, up to the order of $\bar k^3$, enforces a stringent constraint on the parameter $\alpha$, namely, $\alpha = 1$. The choice of $\alpha = 1$ corresponds precisely to the situation maintaining the asymptotic behavior of LQC, as described in Ref. \cite{Renevey:2021tmh}.

Furthermore, the modified dispersion relation (Eq. \eqref{QDR}) incorporates an effective mass squared term ${m_{\rm{G}}}^2$ defined in Eq. (\ref{mG}) with Eq. (\ref{TQ}). Notably, this ${m_{\rm{G}}}^2$ term, governed by the yet undetermined function $g\left( {\bar k,\bar p} \right)$, lacks a guarantee of strict positivity. The stipulation for ${m_{\rm{G}}}^2$ to be positive indeed imparts an additional constraint on the form of $g\left( {\bar k,\bar p} \right)$. Upon substituting Eq. (\ref{SFog}) and Eq. (\ref{fX-1}), we arrive at the expression:
\begin{align}
\label{mG2-1} %"the 1st version of the expression of m_G^2"
{m_{\rm{G}}}^2 &=  - \frac{5}{2}C\left( {1 - \alpha } \right)\frac{{\gamma \bar \mu {{\bar k}^3}}}{{\bar p}} + \frac{1}{{16}}\left[ {44 + 66C - {C^2}\left( {47 - 160\alpha  + 113{\alpha ^2}} \right)} \right]\frac{{{\gamma ^2}{{\bar \mu }^2}{{\bar k}^4}}}{{\bar p}} + {\cal O}\left( {{{\bar k}^5}} \right) \nonumber\\
& = \frac{11}{{8}}\left( {2 + 3C} \right)\frac{{{\gamma ^2}{{\bar \mu }^2}{{\bar k}^4}}}{{\bar p}} + {\cal O}\left( {{{\bar k}^5}} \right) \;,
\end{align}
where the second equality is obtained by imposing the condition $\alpha = 1$. This indicates that ${m_{\rm{G}}}^2$ is bound to be positive on the premise of setting $\alpha$ as unity. In other words, the requirement of a positive ${m_{\rm{G}}}^2$ term does not introduce any supplementary constraints on the free parameters, thereby leaving the parameter $C$ unconstrained within our considerations. Moreover, it is evident that ${m_{\rm{G}}}^2$ exhibits a linear increase with the parameter $C$, up to the order of $\bar k^4$.

%----------------------------------------------------
\section{Constraints on the effective mass of the graviton}
\label{4} %"the 4th section of the main body"
%----------------------------------------------------
In this section, we will further delve into a detailed exploration of the effective mass of the graviton when considering the specific form \eqref{SFog} with Eq. \eqref{fX-1} of the generalized holonomy corrections. Our emphasis will be on scrutinizing the constraints imposed on the effective mass through astronomical observations.

As a preliminary, we first derive the modified Friedman equation with the generalized holonomy corrections. To achieve this, we express the effective total Hamiltonian constraint for the homogeneous and isotropic background in the following manner:
\begin{equation}
\label{QTBHC} %"Quantum corrected Total Background Hamiltonian Constraint"
\bar{H}^{\rm{Q}}\left [ \bar{N} \right ] = \bar{H}^{\rm{Q}}_{\rm{G}}\left [ \bar{N} \right ] +\bar{H}_{\rm{M}}\left [ \bar{N} \right ] = \frac{\bar{N} V_{0}}{{16{\rm{\pi }}}}\left( - 6\sqrt {\bar p} \frac{{{f^2}\left( {\gamma \bar \mu \bar k} \right)}}{{\gamma^2 \bar \mu^2 }} \right) + \bar{H}_{\rm{M}}\left [ \bar{N} \right ] \;,
\end{equation}
where the function $f$ given in Eq. \eqref{fX-1} now has been applied with the restriction condition $\alpha = 1$ from anomaly cancellation and becomes:
\begin{equation}
\label{fX-2} %"the 2nd version of the expression of f(X)"
f\left( X \right) = \sqrt {\frac{{1 - {{\cos }^{2\left( {C + 1} \right)}}\left( X \right)}}{{C + 1}}}\;.
\end{equation}
Then varying this Hamiltonian constraint with respect to the background lapse $\bar{N}$ gives rise to the Hamiltonian constraint equation, i.e. the effective Friedmann equation:
\begin{equation}
\label{HCE} %"Hamiltonian Constraint Equation"
0 = \frac{\delta \bar{H}^{\rm{Q}}\left [ \bar{N} \right ]}{\delta \bar{N}} = -\frac{3 \sqrt{\bar p}}{{8{\rm{\pi }}}} \frac{{{f^2}\left( {\gamma \bar \mu \bar k} \right)}}{{\gamma^2 \bar \mu^2 }} + \frac{\delta \bar{H}_{\rm{M}}\left [ \bar{N} \right ]}{\delta \bar{N}} = \bar{p}^{3/2} \left ( -\frac{3}{{8{\rm{\pi }}}} \frac{{{f^2}\left( {\gamma \bar \mu \bar k} \right)}}{{\gamma^2 \Delta}} + \bar{\rho} \right ) \;,
\end{equation}
where $\bar{\rho}$ is the background energy density defined as $\bar{\rho} := \frac{1}{\bar{p}^{3/2}} \frac{\delta \bar{H}_{\rm{M}}\left [ \bar{N} \right ]}{\delta \bar{N}}$. The above effective Friedmann equation can be further reduced to:
\begin{equation}
\label{EFE-1} %"the 1st version of Effective Friedmann Equation"
{{f^2}\left( {\gamma \bar \mu \bar k} \right)} = \frac{\bar{\rho}}{\rho_{\rm{c}}} \;,
\end{equation}
or
\begin{equation}
\label{EFE-2} %"the 2nd version of Effective Friedmann Equation"
{1 - {{\cos }^{2\left( {C + 1} \right)}}\left( {\gamma \bar \mu \bar k} \right)} = \frac{\bar{\rho}}{\rho_{\rm{c}}/\left ( {C + 1} \right )} \;.
\end{equation}
 In the given equation, we introduce a constant $\rho_{\rm{c}} := \frac{3}{{8{\rm{\pi }} \gamma^2 \Delta}}$, representing the critical energy density at which the bounce happens in conventional holonomy-correction LQC \cite{Ashtekar:2006rx, Ashtekar:2006uz, Ashtekar:2006wn}. From Eq. \eqref{EFE-2}, it becomes evident that the energy density $\bar{\rho}$ exhibits a new finite upper critical value of ${\rho_{\rm{c}}/\left ( {C + 1} \right )}$, decreasing as $C$ increases. This contrasts with the conventional holonomy-correction scenario, where the upper bound of the energy density $\bar{\rho}$ remains the definite constant ${\rho_{\rm{c}}}$. For more discussions on this effective Friedmann equation with generalized holonomy corrections, readers can also refer to Ref. \cite{Renevey:2021tmh}.

Now, let's delve into the discussion of the effective mass ${m_{\rm{G}}}$ associated with tensor mode perturbations. By combining Eq. \eqref{mG2-1} with Eq. \eqref{EFE-1}, we derive the following expression:
\begin{equation}
\label{mG2-2} %"the 2nd version of the expression of m_G^2"
{m_{\rm{G}}}^2 \simeq \frac{11}{{8}}\left( {2 + 3C} \right)\frac{1}{\gamma ^2 \Delta}\left ( \frac{\bar{\rho}}{\rho_{\rm{c}}} \right )^2 \;.
\end{equation}
The equation above indicates that the effective mass ${m_{\rm{G}}}$ depends on the background energy density $\bar{\rho}$, which evolves as the universe expands. Consequently, ${m_{\rm{G}}}$ evolves with time. Nevertheless, this temporal evolution can be ignored because we only focus on the low-curvature phase.
We would like to emphasize once more that ${m_{\rm{G}}}^2$ exhibits nearly linear growth with respect to the parameter $C$.

Then, we can estimate the value of the effective mass ${m_{\rm{G}}}$ of the graviton at the present epoch according to Eq. \eqref{mG2-2}. Given the orders of magnitude of $\gamma \sim {\cal O}\left( 1 \right)$, $\Delta \sim {\cal O}\left( 1 \right)$, ${\rho_{\rm{c}}} \sim {\cal O}\left( 1 \right)$ and the energy density of the present universe $\bar{\rho}(t_0)\sim {\cal O}\left( 10^{-120} \right)$, one obtains that ${m_{\rm{G}}}(t_0)\sim {\cal O}\left( \sqrt{C} \right){\cal O}\left( 10^{-120} \right)$, which corresponds to ${m_{\rm{G}}}(t_0) \sim {\cal O}\left( \sqrt{C} \right){\cal O}\left( 10^{-92} \right){{\rm{eV}}/c^2}$ 
after filling in the physical constants $c$, $G$ and $\hbar$. Latest observational data of pulsar timing arrays (PTAs) and their analysis have established the state-of-the-art upper bound on the graviton mass of ${m_{\rm{G}}} < {\cal O}\left( 10^{-24} \right){{\rm{eV}}/c^2}$ \cite{Wang:2023div, Wu:2023rib}. And this upper bound could continue to be lowered in future gravitational wave measurements. Consequently, the parameter $C$ obtains a conservative constraint of $C<{\cal O}\left( 10^{136} \right)$. This constraint appears to be too loose, and we believe it is mainly limited by the accuracy of the observational upper bound of the graviton mass. Unfortunately, due to the extremely large order of magnitude of the upper bound on $C$ obtained at present, it is unlikely to obtain an adequately tight restriction by observations in the near future. On the other hand, provided that the parameter $C\sim {\cal O}\left( 1 \right)$, our theoretically estimated effective graviton mass ${m_{\rm{G}}}(t_0)$ will be far below the observational upper bound currently. Nonetheless, considering the dependence of the effective graviton mass ${m_{\rm{G}}}$ on the background energy density $\bar{\rho}$, this effective mass might play a significant role in some physical phenomena in the early universe, such as inflation.

%----------------------------------------------------
%----------------------------------------------------
\section{Summary and outlook}
\label{conclusion}
%----------------------------------------------------
In this paper, we study the quantum-corrected dynamics of the tensor mode in a flat cosmological background. To take into account quantum ambiguities inherent in holonomy corrections in LQC, we introduce a generalized function $g\left( {\bar k,\bar p} \right)$ as a replacement for the classical background connection variable $\bar k$ in the Hamiltonian constraint. This is in contrast to the conventional form $\frac{{\sin \left(\gamma \bar \mu \bar k\right)}}{{\gamma \bar \mu }}$ used in the standard $\bar{\mu}$-scheme. Starting from this effective Hamiltonian with generalized holonomy corrections, we derive the modified tensor mode equation and the corresponding dispersion relation for GW propagation within the Hamiltonian framework. The general form of the quantum-corrected dispersion relation, which includes an additional term ${m_{\rm{G}}}^2$ compared to the classical one, remains independent of the specific functional form of $g\left( {\bar k,\bar p} \right)$. However, the specific expression for this effective mass ${m_{\rm{G}}}$ depends on the choice of the function $g\left( {\bar k,\bar p} \right)$. 
What's more, the requirement of a positive ${m_{\rm{G}}}^2$ term prevents causality violations, and then the different proper wave number $k/a$ modes of the GWs propagate with different group velocities $v_{{\rm{GW}}}^{\rm{Q}}$ that less than the speed of light.

Considering the automatically anomaly-free constraint algebra for the tensor mode, anomaly cancellation for vector modes is employed to restrict the possible functional form of $g\left( {\bar k,\bar p} \right)$. To be more clear and accessible, a specific possible functional form is explored, and then anomaly cancellation up to the leading order of the background extrinsic curvature $\bar k$ gives an explicit restriction on the undetermined parameters. In addition, it is worth mentioning that the requirement of positive ${m_{\rm{G}}}^2$ does not impose any additional restriction on these parameters. However, it is necessary to emphasize that this specific form of $g\left( {\bar k,\bar p} \right)$ meeting the restriction ensures an anomaly-free constraint algebra for vector modes only up to the leading order. Consequently, the significance of anomaly terms becomes apparent at larger values of $\bar k$, particularly in the vicinity of the bounce phase. That is, the applicability of this functional form is primarily limited to the small $\bar k$ phase. Finally, we would like to emphasize that it is the first instance to explore the influence of generalized holonomy corrections on the dispersion relation of tensor perturbations. In particular, leveraging the modified dispersion relation, we study the constraints imposed on the effective mass through astronomical observations.

In the future, a crucial objective is to calculate the perturbative constraint algebra for scalar modes with generalized holonomy corrections, all while excluding the incorporation of counter terms. This endeavor holds paramount importance as it could provide valuable insights into the fundamental constraints and symmetries governing the behavior of these modes in the context of LQG from a theoretical perspective. On the other hand, once we have the anomaly-free constraint algebra at our disposal, we can proceed to derive gauge-invariant cosmological perturbation equations for scalar modes with generalized holonomy corrections. Such perturbation equations allow us to probe the imprint of quantum gravity on the CMB, offering a unique opportunity to bridge the gap between theoretical quantum gravity frameworks and observable cosmological phenomena. The specific research plan is outlined as follows:
\begin{itemize}
	\item The first step is to study the impact of quantum gravity effects on the primordial power spectra. Just as mentioned in the introduction, Refs. \cite{Renevey:2021tmh} and \cite{DeSousa:2022rep} have investigated the primordial power spectra with generalized holonomy corrections. However it's essential to note that the effects of these corrections in Ref. \cite{Renevey:2021tmh} only manifest at the background level. While the study in Ref. \cite{DeSousa:2022rep} considers corrections at both the background and perturbation levels, the incorporation of counter terms within the constraints has the potential to introduce quantum gravity corrections beyond holonomy corrections, potentially yielding unexpected corrections.
	In our future plans, we aim to pursue a theoretically consistent gauge-invariant cosmological theory at both the background and perturbative levels, excluding the inclusion of counter terms. We anticipate obtaining intriguing and observable results that differ from those explored in Refs. \cite{Renevey:2021tmh,DeSousa:2022rep}.
	\item Furthermore, we aim to conduct a thorough analysis of the angular power spectra, incorporating generalized holonomy corrections. This approach will allow us to explore the imprint of quantum gravity on the temperature fluctuations across different angles on the CMB.
	\item We can also explore the quantum gravity effects in the GW background induced by the linear cosmological scalar perturbations, incorporating generalized holonomy corrections during the radiation-dominated phase of the early universe. Our objective is to conduct a comprehensive analysis of the angular power spectrum associated with these scalar-induced GWs, aiming to unveil distinct imprints of quantum gravity.
\end{itemize}
We would like to emphasize that, as we all known, there is currently a lack of comprehensive analysis on the quantum gravity effects in scalar-induced gravitational waves, even only incorporating the usual holonomy corrections. In addition, a complete analysis of the angular power spectra is also absent. We aspire to address this gap and complete the puzzle. This effort represents a unique opportunity to bridge the gap between theoretical quantum gravity frameworks and observable cosmological phenomena. Work on these endeavors is currently in progress.

%----------------------------------------------------
%----------------------------------------------------
\acknowledgments
We are very grateful to Martin Bojowald, Ruo-Ting Chen, Yu Han,  Tianyu Jia, Yen Chin Ong, Yu Sang and Ruihong Yue for helpful discussions and suggestions. This work is supported by National Key R$\&$D Program of China (No. 2020YFC2201400), the Natural Science Foundation of China under Grants No. 12375055, and the Top Talent Support Program from Yangzhou University.

%----------------------------------------------------
\bibliographystyle{utphys}
\bibliography{Ref}
\end{document}